\documentstyle[11pt,epsfig]{article}
\newcommand{\ba}{\begin{array}}
\newcommand{\ea}{\end{array}}
\newcommand{\bd}{\begin{displaymath}}
\newcommand{\ed}{\end{displaymath}}
\newcommand{\be}{\begin{equation}}
\newcommand{\ee}{\end{equation}}
\newcommand{\bea}{\begin{eqnarray}}
\newcommand{\eea}{\end{eqnarray}}

\def\bra{\langle}
\def\ket{\rangle}

\def\L{\Lambda}

\def\sgn{\rm{sgn}}

\def\BR{\rm{BR}}

\def\q2 {q^2}

 \def\N10{\widetilde \chi_1^0}
                         \def\C1p{\widetilde \chi_1^+}
                         \def\C1m{\widetilde \chi_1^-}
                         \def\C1pm{\widetilde \chi_1^\pm}

\def\beq{\begin{eqnarray}}
\def\enq{\end{eqnarray}}

\def\lsim{\:\raisebox{-0.5ex}{$\stackrel{\textstyle<}{\sim}$}\:}
\def\gsim{\:\raisebox{-0.5ex}{$\stackrel{\textstyle>}{\sim}$}\:}
\def\neut {{\widetilde\chi}^0}
\begin{document}
 \begin{flushright}
{\large HEP-IFUAP-02-03}\\
{\large OITS-729}\\
{\large SHEP-03-05}\\
\end{flushright}
\begin{flushleft}
\end{flushleft}
\begin{center}
{\Large\bf The di-photon signature of Higgs bosons}\\[5mm]
{\Large\bf in GMSB models at the LHC}\\[15mm]
{\large J. Lorenzo Diaz-Cruz$^{a}$\footnote{ldiaz@sirio.ifuap.buap.mx},
Dilip Kumar Ghosh$^{b}$\footnote{dghosh@physics.uoregon.edu}
and Stefano Moretti$^{c}$\footnote{stefano@hep.phys.soton.ac.uk}}\\[4mm]
{\em $^{a}$ Instituto de Fisica, BUAP, Puebla 72570, Mexico}\\[2mm]
{\em $^{b}$ Institute of Theoretical Science, University of Oregon, 
Eugene OR 97403-5203, USA}\\[2mm]
{\em $^{c}$ Department of Physics \& Astronomy, University of
Southampton, \\
Highfield, Southampton SO17 1BJ, UK}\\[10mm]
\end{center}
\begin{abstract}
\noindent
We show how the well studied $\gamma\gamma$ inclusive Higgs signal can be
used at the Large Hadron Collider
to test Gauge Mediated Supersymmetry Breaking
scenarios in which a rather heavy Higgs
boson decays into two light neutralinos, the latter yielding
two photons and missing (transverse) energy.

\end{abstract}

\vskip 1 true cm

\noindent

PACS Nos.: 13.85.-t, 12.60.Fr, 12.60.Jv, 14.70.Bh  

\newpage

\section{Introduction}
\label{Sect:Intro}

The $\gamma\gamma$ (or `di-photon')
inclusive signature is possibly the most studied
one experimentally, in the context of the Large Hadron Collider (LHC), 
since it allows for the detection of a relatively light ($\lsim130$ GeV) 
neutral Higgs boson, both within the Standard Model (SM) and the Minimal
Supersymmetric Standard Model (MSSM) \cite{ATLAS,CMS}. In fact, the existence 
of such a light particle state has been hinted already by LEP2 data,
both as a possible resonance in the region $M_{\rm{Higgs}}\approx115$ GeV
and as the mass range that best accommodates the higher order Higgs boson
contributions to precision Electroweak (EW) data (see, e.g., 
\cite{LEPTRE}, for a review).

Despite having a Branching Ratio (BR) that is only at the permille
level, the di-photon signature is preferred to the one associated with the
main Higgs decay channel in the above mass region, i.e., Higgs~$\to
b\bar b$ (into pair of $b$-quark jets, with BR practically one), 
as the latter is swamped
by the huge QCD background typical of a hadronic machine, whereas
the former is much cleaner in such an environment. Besides, 
the higher precision
that one can achieve in determining both directions and energies of the 
photons (as compared to those of jets), allows one to obtain a Higgs
mass resolution of the order 2--3 GeV, which compares rather well to
a typical 15--20 GeV accuracy from jet events, also recalling that
the Higgs width is at most a few tens of MeV in the above mass region
(so that, the worse the mass resolution the larger the background,
while the signal size remains relatively stable) \cite{LesHouches2001}.  

The di-photon signature may also represent a distinctive 
signature of broken Supersymmetry (SUSY), namely, in Gauge Mediated
Supersymmetry Breaking (GMSB) models \cite{GMSB}. In these scenarios, the
Lightest SUSY Particle (LSP) is the so-called gravitino, $\widetilde G$.
Moreover, the next-to-LSPs (NLSPs) are usually the lightest neutralino,
${\widetilde\chi}^0_{1}$, or the lighter stau, $ \widetilde t_1 $, 
depending upon the actual configuration of the GMSB model. In the first case
then, the following decay chain could well occur:
Higgs~$\to {\widetilde\chi}^0_{i} {\widetilde\chi}^0_{j}\to
\gamma\gamma~+~E_{\rm miss}$ ($i,j=1,2$), where the missing (transverse) 
energy is due to the two LSPs and possibly neutrinos 
(from $\neut_2 \to \neut_1 + \nu \bar \nu $) escaping 
detection\footnote{Notice that, within the MSSM, 
one can obtain the di-photon signature (including $E_{\rm{miss}}$)
of heavy Higgs bosons also from
non-GMSB scenarios in which the LSP is the lightest neutralino,
 $\widetilde \chi^0_1$, such as in (minimal) Supergravity  (mSUGRA)
models. In fact, the following decay chain can occur here: 
$ H/A \to \widetilde \chi^0_2\widetilde \chi^0_2 $ followed
by a double radiative decay  
$\widetilde\chi^0_2\to \widetilde \chi^0_1 \gamma $~\cite{haber}. 
It has been shown in Ref.~\cite{mele} that this channel
can be large if $\tan\beta$ (see below for its definition) is small
($\simeq1$), with or without gaugino mass unification at high scales. 
However, this MSSM configuration  is ruled out by the LEP2 limit 
on the lighter chargino mass \cite{W1LEP2}. Only under the assumption 
of non-universal gaugino masses can the 
BR$(\widetilde\chi^0_2\to \widetilde 
\chi^0_1 \gamma)$ be large at high $\tan\beta$ in
mSUGRA models \cite{Krupo}.}. 

Current collider limits however forbid the lightest MSSM Higgs boson, $h$,
to decay into two neutralinos, i.e., $M_h< 
M_{{\widetilde\chi}^0_{i}}+M_{{\widetilde\chi}^0_{j}}$ ($i,j=1,2$),
so that only the two heavier neutral Higgs bosons, $H,A$,
can initiate the above SUSY decay chain. Recalling that $h\to\gamma\gamma$
direct decays can still occur in GMSB models, it is intriguing to consider
the possibility that {\sl all} neutral Higgs bosons of the MSSM can
be detected in the same channel, that is, a pair of photons accompanied
by some amount of missing (transverse) energy,  $E_{\rm miss}$. 
Notice that the latter should in average be larger
for signal events as compared to the background, which is
dominated by prompt di-photon production, where
$E_{\rm miss}$ is mainly due to jet energy losses down the beam pipe
or to non-fully hermetic detectors. In contrast, one would naturally
expect a large  $E_{\rm miss}$ value arising from the above  
$H$ and $A$ decays
(if not from invisible decay products of heavy particles produced
in association with any of the Higgs states, see below), 
so that the missing (transverse) energy may be used
in the kinematical selection. In practice though, only 
two resonances could in the end be visible, as the $H$ and $A$
bosons are almost degenerate in mass. Besides, 
the latter would tend to be located
at higher $\gamma\gamma$ invariant masses as compared to the one
due to $h\to\gamma\gamma$ decays,
where the production cross section is smaller, but so is the di-photon
continuum.

Finally, whereas the $h\to\gamma\gamma$ resonance can directly be reconstructed
from the photon four-momenta, the same is not true in the
$H,A\to {\widetilde\chi}^0_{i}{\widetilde\chi}^0_{j}\to
\gamma\gamma~+~E_{\rm miss}$ ($i,j=1,2$) channel. Here, however,
after ensuring that the two photons are not back-to-back, one can 
attempt to resolve the 
$E_{\rm miss}$ along their directions and add it to the photonic
transverse momenta,  $p_{T}^{\gamma_1}$ and $p_{T}^{\gamma_2}$. 
Scaling up the respective sums by the ratios $p^{\gamma_1}/p_{T}^{\gamma_1}$ 
and $p^{\gamma_2}/p_{T}^{\gamma_2}$ gives in principle
 the reconstructed momenta of the 
neutralino pair. In practice though, the presence of several
unresolved sources of missing (transverse) energy may spoil the
mass reconstruction.

It is the purpose of this paper to investigate in detail such a phenomenology,
in the context of GMSB scenarios.
After a brief discussion of the parameters defining GMSB models and a
description of the tools used in order to carry out our numerical 
studies, we will present the results and draw our conclusions.
\section{The spectrum in GMSB models}
\label{Sect:GMSB}

In GMSB models, the symmetry of the Superpotential is
broken at some relatively low scale,
say, a few hundred TeV (the `hidden sector'),
and SUSY-breaking 
is communicated to the detectable particles (the `visible sector')   
through so-called `messenger' fields, effectively, gauge bosons.

In fact, 
renormalisability of the theory, coupled with economy of field content,
dictates that the messenger sector (MS) 
be comprised of chiral Superfields such that their SM gauge 
couplings are vectorial in nature. Most GMSB models actually consider 
these fields to be in 
($5 + \bar 5$) or ($10 + \overline{10}$) representations of SU(5). They
are also chosen to transform as a multiplet of a Grand Unification Theory
(GUT),  so that the SUSY prediction
of gauge coupling unification is preserved. These facts restrict the
 maximum number of messenger families $N_M$ to be  
$\leq 4$ and $\leq 1 $ for the ($5 + \bar 5$) and
$(10 + \bar 10)$ constructs, respectively.  

Limiting ourselves, for the time being, to a single pair of 
MS Supermultiplets ($\Psi + \bar \Psi$), consider a term in the 
Superpotential of the form $\lambda {\cal S} \bar \Psi \Psi$, where 
${\cal S}$ is a SM singlet. The scalar ($S$) and auxiliary ($F_S$)
components of ${\cal S}$ may acquire
Vacuum Expectation Values (VEVs) 
through their interactions with the hidden sector fields. SUSY-breaking 
is thus communicated to the MS, with the fermions and sfermions 
acquiring different masses. This, in turn, is communicated to the 
SM fields resulting in the gauginos and sfermions acquiring masses 
at the one-loop and two-loop levels, respectively. The expressions,  
in the general case of multiple messenger pairs and/or gauge singlets
${\cal S}_i$, is a somewhat 
complicated function \cite{GMSB} of $M \equiv \bra S \ket$ and 
$\L \equiv \bra F_S \ket / \bra S \ket$. However, if there is just one such 
singlet, the expressions for soft SUSY-breaking gaugino and scalar 
masses at the messenger scale $M$ simplify to:
\beq
{\tilde M}_i(M) &=& N_M \frac{\alpha_i(M)}{4\pi} 
g_1 \left(\frac{\L}{M}\right)\Lambda,  \\ 
{\tilde m}^2_{\tilde f}(M) &=&  
        2 N_M \L^2  g_2 \left(\frac{\L}{M} \right) 
              \sum^3_{i=1}\kappa_i C^{\tilde f}_i 
                        \left(\frac{\alpha_i(M)}{4\pi}\right)^2.
        \label{threshold}
\enq
In eq. (\ref{threshold}), $ C^{\tilde f}_i$ are the quadratic Casimirs 
for the sfermion in question. The factors $\kappa_i$ equal 1, 1 and 
$5/3$ for SU(3), SU(2) and U(1), respectively, with the 
gauge couplings so normalised that all $\kappa_i \alpha_i$'s are equal 
at the messenger scale. The threshold functions are given by
\beq
g_1(x) &=& \frac{1+x}{x^2} \log(1+x)+ \big (x \to -x \big), \\ 
g_2(x) &=& \frac{(1+x)}{x^2}
        \bigg [\log(1+x) + 2 {\rm{Li}}_2\bigg (\frac{x}{1+x}\bigg ) 
                         - \frac{1}{2}{\rm{Li}}_2\bigg (\frac{2x}{1+x}\bigg )
        \bigg ] + \big (x \to -x \big).
\enq
The Superparticle masses at the EW scale are obtained 
from those in eqs.~(1)--(\ref{threshold}) by evolving the appropriate
Renormalisation Group Equations (RGEs). For the scalar masses, the $D$-terms 
need to be added too. The Higgs 
sector of the `minimal' 
GMSB model contains the two usual Higgs doublets
of the MSSM,  $(H_u, H_d)$. 
The ratio of the VEVs of the latter is parameterised as
$\tan\beta =\frac{v_u}{v_d}$. Moreover, in
the Superpotential one has a  Higgs bilinear term of the form
 $B\mu H_u H_d$. In general, $\mu$ and $B$ depend
on the details of the SUSY-breaking mechanism. However, if we 
assume that the EW symmetry is broken radiatively, then the values of
$\mu^2$ and $B$
are determined in terms of the other parameters of the model. Without
loss of generality, one may express the entire particle spectrum of 
such a  GMSB model in terms of five external inputs only:
$M,\Lambda, \tan\beta$, sgn$(\mu)$ and $N_M$ (hereafter, 
we assume $N_M=1$).

\section{Parameter scans}
\label{Sect:Scans}

As a starting point of our investigation, we shall discuss the
relevant particle spectrum by searching for regions of the GMSB 
parameter space where the heavy Higgs bosons,
$H$ and $A$, and the lightest neutralino, ${\widetilde\chi}^0_{1}$,
have masses and compositions (in terms of the gaugino-Higgsino fields)
such that the decays 
$H,A \to {\widetilde\chi}^0_{1} {\widetilde\chi}^0_{1}$ are kinematically
allowed and reach BRs that are at least comparable to that of the
SM Higgs decay into two photons, for which one has
BR$(H_{\rm{SM}} \to \gamma\gamma) \simeq 10^{-3}$. In fact, such
a channel is of extreme importance in the MSSM too, as already noted. 
Here,
the parameters $\tan\beta$ and $M_A$ define entirely the Higgs sector of 
the MSSM
at tree-level, $\tan\beta$ having being already defined 
and with $M_A$ being the mass of the pseudoscalar
Higgs state $A$ (the other two states, $h$ and $H$, are scalars)\footnote{In 
fact,
in the GMSB model, all Higgs masses are derived quantities, as 
specified in the previous Section.}. It turns out that,
for any $\tan\beta$ value, if $M_A\gsim 150$--200 GeV,
the $h\to\gamma\gamma$ decay mode is a discovery channel of the
lightest Higgs boson of the MSSM.

In order to sample the strength of the di-photon signal of our interest,
we present a set of numerical results that include the Higgs and
lightest neutralino masses, as well as the BRs of the channels
$H,A \to {\neut}_1 {\neut}_1$. In computing these rates we
have used the subroutines of ISAJET 7.58 \cite{ISAJET} that 
implement the GMSB model, with several choices of parameter inputs.
Namely, in Tabs. 1,2 and 3 we have fixed $M=150$ TeV
and taken $\Lambda=75, 100$ and $125$ TeV, respectively.
Here, the sign of $\mu$ is always negative whereas $\tan\beta$ varies from 5 up
to a maximum value where the decays $H,A \to {\neut}_1 {\neut}_1$
are no longer kinematically allowed.

From Tabs. 1--3 one can appreciate the following trends.
\begin{itemize}
\item As intimated in the Introduction, the Higgs masses $M_H$ and $M_A$ 
show a degeneracy 
within a couple of GeVs. 
Thus, we can add their corresponding two-photon signals,
as it would not be possible to distinguish among them
solely on the basis of the reconstructed mass.
\item  The lightest neutralino ${\neut}_1$
decays into a photon plus a gravitino,
${\neut}_1\to \gamma+\widetilde{G}$,
 while BR$(H,A \to {\neut}_1 \neut_1)$ can exceed 
the $10^{-3}$ level, this yielding altogether a decay rate for the
channels $H,A\to {\widetilde\chi}^0_{1}{\widetilde\chi}^0_{1}\to
\gamma\gamma~+~E_{\rm miss}$  above our reference di-photon SM Higgs decay 
rate.
\item For values of $\tan\beta$ larger than 40 or so, the decays
$H,A \to \neut_1 \neut_1 $ are no longer kinematically allowed.
\end{itemize}

\noindent
{Table~1: 
Higgs, lightest neutralino masses and BR$(H,A \to {\neut_1} {\neut_1})$ 
for a sample set of GMSB inputs with $(M,\Lambda)=(150,75)$ \,TeV. }
\vspace*{1.5mm}
\begin{center}
\begin{tabular}{|c|c|c|c|c|c|}
\hline
$\tan\beta$ & $M_H$ [GeV] & $M_A$ [GeV] & $M_{\neut_1}$ [GeV] &
BR$(H)$ & BR$(A)$ \\
\hline
 5  &  430 & 428 & 105 & $1.4\times 10^{-2}$ & $1.1\times 10^{-2}$ \\
\hline
10  &  404 & 403 & 104 & $1.1\times 10^{-2}$ & $1.2\times 10^{-2}$ \\
\hline
15  &  392 & 391 & 103 & $6.0\times 10^{-3}$ & $7.3\times 10^{-3}$ \\
\hline
20  &  377 & 377 & 103 & $3.5\times 10^{-3}$ & $4.6\times 10^{-3}$ \\
\hline
25  &  358 & 358 & 103 & $2.1\times 10^{-3}$ & $3.0\times 10^{-3}$ \\
\hline
30  &  332 & 332 & 103 & $1.3\times 10^{-3}$ & $2.0\times 10^{-3}$ \\
\hline
35  &  296 & 296 & 103 & $7.5\times 10^{-4}$ & $1.4\times 10^{-3}$ \\
\hline
40  &  245 & 245 & 102 & $2.5\times 10^{-4}$ & $7.9\times 10^{-4}$ \\
\hline
45  &  162 & 161 & 102 &        0            &   0                 \\
\hline
\end{tabular}
\end{center}

\newpage

\noindent
{Table~2: 
Higgs, lightest neutralino masses and BR$(H,A \to {\neut_1} {\neut_1})$ 
for a sample set of GMSB inputs with $(M,\Lambda)=(150,100)$ \,TeV. }
\vspace*{1.5mm}
\begin{center}
\begin{tabular}{|c|c|c|c|c|c|}
\hline
$\tan\beta$ & $M_H$ [GeV] & $M_A$ [GeV] & $M_{\neut_1}$ [GeV] &
BR$(H)$ & BR$(A)$ \\
\hline
 5  &  552 & 551 & 146 & $7.6\times 10^{-3}$ & $7.5\times 10^{-3}$ \\
\hline
10  &  521 & 520 & 145 & $7.1\times 10^{-3}$ & $8.6\times 10^{-3}$ \\
\hline
15  &  505 & 505 & 145 & $3.7\times 10^{-3}$ & $4.9\times 10^{-3}$ \\
\hline
20  &  488 & 487 & 144 & $2.1\times 10^{-3}$ & $3.0\times 10^{-3}$ \\
\hline
25  &  464 & 464 & 144 & $1.2\times 10^{-3}$ & $1.9\times 10^{-3}$ \\
\hline
30  &  433 & 432 & 144 & $7.6\times 10^{-4}$ & $1.3\times 10^{-3}$ \\
\hline
35  &  390 & 389 & 144 & $4.1\times 10^{-4}$ & $8.6\times 10^{-4}$ \\
\hline
40  &  330 & 330 & 144 & $1.2\times 10^{-4}$ & $4.7\times 10^{-4}$ \\
\hline
45  &  237 & 237 & 144 &        0            &   0                 \\
\hline
\end{tabular}
\end{center}

\bigskip

\noindent
{Table~3: 
Higgs, lightest neutralino masses and  BR$(H,A \to {\neut_1} {\neut_1})$ 
for a sample set of GMSB inputs with $(M,\Lambda)=(150,125)$ \,TeV. }
\vspace*{1.5mm}
\begin{center}
\begin{tabular}{|c|c|c|c|c|c|}
\hline
$\tan\beta$ & $M_H$ [GeV] & $M_A$ [GeV] & $M_{\neut_1}$ [GeV] &
BR$(H)$ & BR$(A)$ \\
\hline
 5  &  666 & 664 & 197 & $4.8\times 10^{-3}$ & $5.4\times 10^{-3}$ \\
\hline
10  &  628 & 628 & 195 & $4.8\times 10^{-3}$ & $6.5\times 10^{-3}$ \\
\hline
15  &  610 & 610 & 195 & $2.5\times 10^{-3}$ & $3.7\times 10^{-3}$ \\
\hline
20  &  590 & 589 & 195 & $1.4\times 10^{-3}$ & $2.2\times 10^{-3}$ \\
\hline
25  &  562 & 562 & 195 & $7.9\times 10^{-4}$ & $1.4\times 10^{-3}$ \\
\hline
30  &  526 & 526 & 195 & $4.4\times 10^{-4}$ & $9.3\times 10^{-4}$ \\
\hline
35  &  477 & 476 & 195 & $2.1\times 10^{-4}$ & $5.9\times 10^{-4}$ \\
\hline
40  &  408 & 408 & 194 & $2.3\times 10^{-5}$ & $2.3\times 10^{-4}$ \\
\hline
45  &  306 & 305 & 194 &        0            &   0                 \\
\hline
\end{tabular}
\end{center}

\bigskip

For some regions of the GMSB parameter space, the decays 
$H, A\to \neut_1\neut_2,
\neut_2\neut_2 $, followed by $\neut_2 \to \neut_1 + X$ can be significant
enough to contribute to the two-photon signals. Although this effect was not 
included in the previous Tables, as it was small,
it will be considered in the remainder of this Section
and in the Monte Carlo (MC) simulations of the next one as well.
To discuss their effects,  we have produced three additional
sample points, which we elaborate upon below and in
Tabs. 4--6, before moving on to the event generator analysis. 

Tab. 4 corresponds to the Snowmass slope $M=2\Lambda$ \cite{SPS}, as we
have fixed $\tan\beta=15$ and taken $\sgn(\mu)$ positive, further
varying $\Lambda$ from 75 to 150 TeV (for lower values of
$\Lambda$ we get a chargino lighter than 150 GeV, which is not
allowed by current collider bounds \cite{W1Tevatron}). Both the BR columns
(BR$({H})$ and BR$({A})$) contain three rows, each corresponding from top
to bottom to the decay rates of $H$ and $A$ 
into $\neut_1\neut_1$, $\neut_1\neut_2$ and $\neut_2\neut_2$ 
pairs, respectively. 

For these sets of GMSB parameter space points the BR of $H$ and $A$ 
into $\neut_1\neut_2$ always dominates over the other two channels. This can
be understood in terms of the enhancement of the $H,A-\neut_1-\neut_2 $ 
coupling, which over-compensates the phase-space suppression in the 
$H,A \to \neut_1\neut_2$ decay modes. Furthermore, one can see from  
Tab. 4 that, as $\Lambda$ increases, the Higgs masses $M_H$ and $M_A$
become rather heavy and they still show degeneracy. The discussed
BRs stay above $10^{-3}$, but because the production rate decreases
substantially for Higgs masses above about 600 GeV, only the lower
values of $\Lambda$ ($\simeq 75$ TeV) will produce sizable
signals. It is interesting to note that in this case the BR
of the lightest Higgs boson into two photons remains close
to the SM value, with $M_h\simeq 115$ GeV.

\bigskip

\noindent
{Table~4: 
Higgs, lightest and second lightest neutralino masses and  
BR$(H,A \to {\neut_1} {\neut_1}, {\neut_1} {\neut_2},{\neut_2} {\neut_2})$ 
for a sample set of GMSB inputs with $M=2\Lambda$, 
$\tan\beta=15$ and $\sgn(\mu)$ positive. }
\vspace*{1.5mm}
\begin{center}
\begin{tabular}{|c|c|c|c|c|c|c|}
\hline
$\Lambda$ [TeV] & $M_H$ [GeV] & $M_A$ [GeV] & $M_{\neut_1}$ [GeV] & 
  $M_{\neut_2}$ [GeV] & BR$(H)$ & BR$(A)$ \\
\hline
\cline{1-5}
 75  &  395 & 394 & 101 & 184 & $7.5\times 10^{-3}$  & $1.0\times 10^{-2}$ \\
     &      &     &     &     & $2.08\times 10^{-2}$ & $4.4\times 10^{-2}$ \\ 
     &      &     &     &     & $4.01\times 10^{-3}$ & $3.3\times 10^{-2}$ \\
\hline
 80  &  419 & 418 & 108 & 198 & $6.5\times 10^{-3}$ & $9.3\times 10^{-3}$ \\
     &   &  & & & $1.8\times 10^{-2}$ & $4.1\times 10^{-2}$                \\ 
& &   &     &  & $3.05\times 10^{-3}$  & $3.02\times 10^{-2}$         \\
\hline
 90  &  467 & 466 & 123 & 225 & $5.2\times 10^{-3}$ & $7.6\times 10^{-3}$ \\
    &       &     &     &   & $1.56\times 10^{-2}$ & $3.7\times 10^{-2}  $ \\
    &       &     &     &   & $1.56 \times 10^{-3}$ & $2.4\times 10^{-2}  $ \\
\hline
100  &  515 & 514 & 137 & 252 & $4.2\times 10^{-3}$ & $6.2\times 10^{-3}$ \\
     &      &     &     &  & $1.32\times 10^{-2}$ & $ 3.25\times 10^{-2}   $ \\
     &      &     &     &  & $ 6.47\times 10^{-4}$ & $1.74\times 10^{-2}  $ \\
\hline
125  &  631 & 631 & 173 & 320 & $2.8\times 10^{-3}$ & $4.3\times 10^{-3}$ \\
     &      &     &     &     & $9.2 \times 10^{-3} $ & $2.55\times 10^{-2}$\\
     &      &     &     &     &   -   &  - \\ 
\hline
150  &  745 & 744 & 209 & 387 & $2.0\times 10^{-3}$ & $3.2\times 10^{-3}$ \\
     &      &     &     &     & $6.9\times 10^{-3} $ & $ 2.1\times 10^{-2}$\\
     &      &     &     &     & - & - \\    
\hline
\end{tabular}
\end{center}

\newpage

\noindent
{Table~5: 
Higgs, lightest and second lightest neutralino masses and  
BR$(H,A \to {\neut_1} {\neut_1}, {\neut_1} {\neut_2},{\neut_2} {\neut_2})$ 
for a sample set of GMSB inputs with $M=2\Lambda$, $\Lambda = 75$ TeV,  
$\tan\beta=10$ and $\sgn(\mu)$ positive. }
\vspace*{1.5mm}
\begin{center}
\begin{tabular}{|c|c|c|c|c|c|c|}
\hline
$\Lambda$ [TeV] & $M_H$ [GeV] & $M_A$ [GeV] & $M_{\neut_1}$ [GeV] & 
  $M_{\neut_2}$ [GeV] & BR$(H)$ & BR$(A)$ \\
\hline
\cline{1-5}
     &  &  &  &  & $1.54\times 10^{-2}$  & $2.0\times 10^{-2}$ \\
 75  & 405 & 404 & 101 & 183 & $4.36\times 10^{-2}$ & $8.5\times 10^{-2}$ \\ 
     &      &     &     &    & $1.2\times 10^{-2}$ & $7.0\times 10^{-2}$ \\
\hline
\end{tabular}
\end{center}

\bigskip 

\noindent
{Table~6: 
Higgs, lightest and second lightest neutralino masses and  
BR$(H,A \to {\neut_1} {\neut_1}, {\neut_1} {\neut_2},{\neut_2} {\neut_2})$ 
for a sample set of GMSB inputs with $M=2\Lambda$, $\Lambda = 75$ TeV,  
$\tan\beta=35$ and $\sgn(\mu)$ negative. }
\vspace*{1.5mm}
\begin{center}
\begin{tabular}{|c|c|c|c|c|c|c|}
\hline
$\Lambda$ [TeV] & $M_H$ [GeV] & $M_A$ [GeV] & $M_{\neut_1}$ [GeV] & 
  $M_{\neut_2}$ [GeV] & BR$(H)$ & BR$(A)$ \\
\hline
\cline{1-5}
     &     &     &     &     & $7.53\times 10^{-4}$  & $1.4\times 10^{-3}$ \\
 75  & 296 & 296 & 102 & 188 & $9.73\times 10^{-5}$ & $2.0\times 10^{-3}$ \\ 
     &      &     &     &    &   -  & -\\
\hline
\end{tabular}
\end{center}

\section{Monte Carlo simulation of signal and background}
\label{Sect:Calculation}
In this section, we evaluate the inclusive 
production cross sections at the LHC 
for the two heavy neutral Higgs bosons of the MSSM, $H$ and $A$, each
followed by all possible decays yielding two photons
and missing (transverse) energy. 

Hereafter, we make the assumption that coloured SUSY particles 
(chiefly, squarks) are heavy enough\footnote{As it is
typical in most GMSB scenarios.}  so that they do not enter 
the loops 
in the `$gg\to $~Higgs' production mode nor they can produce
Higgs bosons in cascade decays or enter the Higgs decay chains
(the same for gluinos).
Alongside the mentioned gluon-fusion channel,
we consider the following Higgs production
modes: `$q\bar q,gg\to Q\bar Q$~Higgs' (associated heavy-quark 
production), 
`$qq\to qq V^*V^*\to qq$~Higgs' (vector-fusion) and 
`$q\bar q^{(')} \to V$~Higgs' (Higgs-strahlung), 
with $q$ representing all possible combinations
of light (anti)quarks and where $Q=b,t$, $V=W^\pm,Z$
and Higgs~$=H,A$ (except in the last two modes, where
the pseudoscalar Higgs boson cannot be produced). These are the 
leading production modes of
neutral Higgs states of the MSSM at the LHC (under the above
assumption of heavy squarks and gluinos)\footnote{See 
Ref.~\cite{production} for a review of their 
properties. } and have been computed
here at next-to-leading order accuracy, by adopting the programs
described at \cite{Spira}, with default settings. 

As for the calculation of the SUSY decays rates 
(again, produced with ISAJET v7.58),
we have adopted the GMSB configurations given in Tab.~7 for $M$ and $\Lambda$,
further varying $\tan\beta $ from $5$ to $40$, always with $\sgn(\mu)> 0$.
(Note that all the above choices of SUSY parameters are allowed by the 
lighter chargino mass limit, as derived within the GMSB model.) 

\bigskip
\noindent
{Table~7: Sets of GMSB parameter points $M$ 
and $\Lambda$ considered in Fig.~1.}
\vspace*{1.5mm}  
\begin{center}
\begin{tabular}{|c|c|c|}
\hline
Point & $M$~[TeV] & $\Lambda $~[TeV] \\
\hline
$A$ & 150 & 75 \\
\hline
$B$ & 150 & 100 \\
\hline
$C$ & 150 & 125 \\
\hline
$D$ & 200 & 75 \\
\hline
$E$ & 200 & 100 \\
\hline
$F$ & 200 & 125 \\
\hline 
$G$ & 200 & 150 \\
\hline
$H$ & 200 & 175 \\
\hline        
\end{tabular}
\end{center}

\bigskip 

We define the `effective' production rate of the 
$\gamma\gamma~+~E_{\rm miss}$ signature, $\sigma_{\rm{eff}}$, as:
\beq\label{sigma}
\sigma_{\rm{eff}} &= &\sigma (gg \to H) \times {\BR}_{\rm{eff}}(H)\nonumber \\
             &+ & \sigma (gg \to A) \times {\BR}_{\rm{eff}}(A)\nonumber \\
             &+ & \sigma (q \bar q, gg \to Q \bar Q H) 
	     \times {\BR}_{\rm{eff}}(H) \nonumber \\
             &+ & \sigma (q \bar q, gg  \to Q \bar Q A) 
	     \times {\BR}_{\rm{eff}}(A) \nonumber\\
             &+ & \sigma (q \bar q \to H V) \times {\BR}_{\rm{eff}}(H) \nonumber \\
             &+ & \sigma (q q \to q q H ) 
	     \times {\BR}_{\rm{eff}}(H) 
\enq
where  
${\BR}_{\rm{eff}}(H)$ and ${\BR}_{\rm{eff}}(A)$ are defined as follows:
\beq
{\BR}_{\rm{eff}}(H/A) &= &\bigg[ {\BR} \bigg (H/A \to \neut_1\neut_1 \bigg ) 
          + {\BR} \bigg (H/A \to \neut_1\neut_2 \bigg ) \times 
	  {\BR} \bigg (\neut_2 \to \neut_1  + \sum_{i=1,...3} \nu_i \bar \nu_i  \bigg )\nonumber\\ 
	  &&+ {\BR} \bigg (H/A \to \neut_2\neut_2 \bigg ) \times 
	  {\BR} \bigg (\neut_2 \to \neut_1  + \sum_{i=1,...3} \nu_i \bar \nu_i \bigg )^2\bigg]
	  \times {\BR}  \bigg (\neut_1 \to \gamma \widetilde G  \bigg )^2.
\enq
\noindent 
In Fig.~1 we display the variation of $\sigma_{\rm {eff}}$ 
with $\tan\beta$ for the sets of GMSB parameter space points given
in Tab. 7. From the pattern of each curve 
it is clear that the lower $\Lambda$ the higher
$\sigma_{\rm{eff}}$. Furthermore, at low $\tan\beta$, $\sigma_{\rm{eff}}$ is larger than
at high $\tan\beta$. This can be understood from the fact that, as
$\tan\beta $ grows, the lighter stau $(\tilde \tau_1 )$ becomes 
the NLSP, with the decay
$\neut_1 \to \tilde \tau_1^\pm + \tau^\mp $ becoming the dominant one:  
this explains the sharp fall in $\sigma_{\rm{eff}}$ at large $\tan\beta$. 

The event simulation has been carried out by exploiting the SUSY
implementation \cite{me} of the {\tt HERWIG} event generator \cite{HERWIG},
supplemented by the ISASUSY \cite{ISAJET} subroutines (v7.58) interfaced
to the {\tt ISAWIG} code \cite{ISAWIG} for SUSY spectra generation. 
We list here the series of process numbers used for the MC event
generation: i.e., 
\begin{verbatim}
                IPROC = 3320 3360 3375 3630 3815 3826
                   3310 3325 3365 3610 3710 3816 3835
                   3315 3335 3370 3620 3720 3825 3836
\end{verbatim}
for the signal and ${\tt IPROC}=2200$ for the background (corresponding
to direct di-photon production $q\bar q,gg\to \gamma\gamma$)\footnote{The 
attentive reader will notice that we have generated in the MC simulations
more processes than those used in Fig.~1. This has been
done for completeness mainly, as the four channels described in 
Sect.~\ref{Sect:Scans} are indeed those making up most of the visible
cross section. Furthermore, all possible decay channels of Higgs bosons and 
neutralinos are included in the MC simulation, through the {\tt ISAWIG} input
files. Also notice that the (inclusive) rates in Fig.~1 use next-to-leading
order normalisation, whereas those in Figs.~2--4 adopt the lowest
order one, as default in a MC event generator.}.

As illustrative values for the MC simulation, we have used the three
points given in Tab. 8. Notice that they have already been discussed,
as they are those appearing in: the fourth 
row of Tab. 4 (Point 1, which is SPS8 of Ref.~\cite{SPS}), 
           Tab. 5 (Point 2) and 
           Tab. 6 (Point 3).

\bigskip 

{Table~8: Sets of GMSB parameter points used in the
{\tt HERWIG} simulation and to produce  Figs.~2--4.}
\vspace*{1.5mm}  
\begin{center}
\begin{tabular}{|c|c|c|c|c|c|}
\hline
Point & $M$~[TeV] &   $\Lambda$~[TeV] &   $N_M$  & $\tan\beta$ & $\sgn(\mu)$ \\
\hline
1 & 200 &      100 &      1 &    15 &  $+$ \\    
2 & 150 &      75 &       1 &    10 &  $+$ \\    
3 & 150 &      75 &       1 &    35 &  $-$ \\     
\hline        
\end{tabular}
\end{center}

\bigskip 

The signature we are trying to extract is simply defined as follows,
along the lines of the ATLAS/CMS triggers \cite{ATLAS,CMS}.
\begin{itemize}
\item Two photons are required:
  one with $p_T > 40$ GeV and 20 GeV for the other, both within 2.5 in 
  pseudorapidity.
\item The two photons are required to have a relative angle less than 175 
  degrees, in order to enable the mass reconstruction of the two heaviest
  Higgs bosons decaying into two photons and two LSPs, the latter yielding
  the missing (transverse) energy.
\item No cuts in missing (transverse) energy are enforced at this preliminary
  stage, nor any restriction on the underlying hadronic activity is imposed.
\end{itemize}
We then look at:
\begin{itemize}
\item $E_{\rm{miss}}$: the missing (transverse) energy.
\end{itemize}
This quantity is plotted in 
Fig.~2(a,b,c), normalised to one. (Hereafter, the labelling a(b)[c] 
in the figures refers to GMSB set 1(2)[3] as given above in Tab.~8.)
As intimated in the Introduction, one may appreciate here the fact that
the missing (transverse) energy distribution is much softer
for the background, as compared to the signal. A suitable cut
on this quantity, which will enhance the signal-to-background ratio,
could be, e.g., $E_{\rm{miss}}>20$ GeV. This may penalise signal
contributions due to $h$ decays, yet it will have
a beneficial impact in extracting those arising from $H$ and $A$.

Upon enforcing this constraint, in addition to the cuts in 1.--2.,
we look at two kinematic observables:
\begin{itemize}
\item $m_{\gamma\gamma}$: the invariant mass of the photon pair obtained by
          using their visible momenta;
\item $M_{\gamma\gamma}$: the invariant mass of the photon pair obtained after 
          resolving the $E_{\rm{miss}}$ along the visible photon directions.
\end{itemize}
These are plotted in Figs.~3(a,b,c) and 4(a,b,c), respectively,
normalised to the integrated cross section in picobarns, as given
by {\tt HERWIG}. From these  plots it is clear the potential
of the LHC in detecting di-photon signals of neutral Higgs bosons,
as induced by our sample of GMSB model configurations. However,
while the direct $h\to\gamma\gamma$ resonance is clearly
visible and narrowly centred around $M_h$
in both distributions  $m_{\gamma\gamma}$ and $M_{\gamma\gamma}$
(and so it was even prior to the enforcement of the cut in
$E_{\rm{miss}}$), no peak associated to $H/A\to \gamma\gamma +
E_{\rm{miss}}$ decays appears, yet
the corresponding events spread over the entire mass range
well above the di-photon background. In fact, our
mass reconstruction procedure fails because of the many unresolvable
sources of missing energy appearing at hadron level, once
all decay  modes of each unstable particle 
are allowed, as dictated by the MSSM configuration induced by the
underlying GMSB scenario.
Here, the exploitation of a more exclusive final state
would be helpful to the cause of extracting the $H/A$ resonance.
However, we refrain here from pursuing this matter further
and simply remark that, even in absence of heavy Higgs mass reconstruction,
a clear excess in the di-photon channel above
the SM expectations should be established at the LHC after
minimal luminosity\footnote{Recall that higher order QCD corrections
to the background are well under control in comparison to the
excesses seen in the last two figures, as they are of order 
10--20\% \cite{gaga}. Moreover, the contribution to the background due to 
$qg\to q\gamma$
with the final state light (anti)quark mistagged as a photon, not
included here, is also small in comparison.}, 
with
some dependence in both the  $m_{\gamma\gamma}$ and $M_{\gamma\gamma}$
spectra upon the relevant masses $M_{H/A}$, $M_{\neut}$ and
$M_{\tilde G}$, whose actual values may be investigated via a comparison
of the data to the MC generated distributions. 

\section{Conclusions}

In summary, we have proved that, for some rather natural configurations
of the GMSB parameter space consistent with current collider
limits, one may be able to extract di-photon signals of all
neutral Higgs bosons of the MSSM at the LHC. While,
after customary ATLAS/CMS 
cuts on the two photons, the mass of the
lightest state would always be visible in the form of a 
$\gamma\gamma$ resonance,
the presence of the two heaviest states (which are degenerate
in mass) can be established, in the form of a clear excess in the total 
number of $\gamma\gamma+E_{\rm{miss}}$ events
over the corresponding
SM predictions, after an additional threshold in missing 
(transverse) energy  is enforced. Thus, after such a selection,
in order to test very specific GMSB model predictions, it suffices
to exploit the total event rate. Finally, the
study of  the kinematics of the entire event sample may allow for
the determination of crucial sparticle masses, 
such as those of the the LSP and NLSP,
hence enabling one to strongly constrain
the underlying SUSY-breaking mechanism.

Our conclusions are based on a sophisticated MC event simulation but a more
rudimentary emulation of detector response (based on
Gaussian smearing of the visible tracks). However, our preliminary
results are rather encouraging and we do believe that 
they call for attention on the ATLAS/CMS side, as the $\gamma\gamma$
channel is possibly the most studied one in the context of Higgs boson
searches at the upcoming CERN hadron collider.

\section*{Acknowledgements}
JLD-C is grateful to the CERN Theory division for hospitality and to
CONACYT-SNI (Mexico) for financial support.
DKG's work is supported by the US DOE contracts DE-FG03-96ER40969 and
DE-FG02-01ER41155. He is also grateful to the Southampton Theory
Group for hospitality while this paper was being completed. The 
authors are grateful to Barbara Mele for illuminating discussions
during the early stages of the analysis.

\begin{figure}[t]
\begin{center}
\hspace{-4.0truecm}\epsfig{file=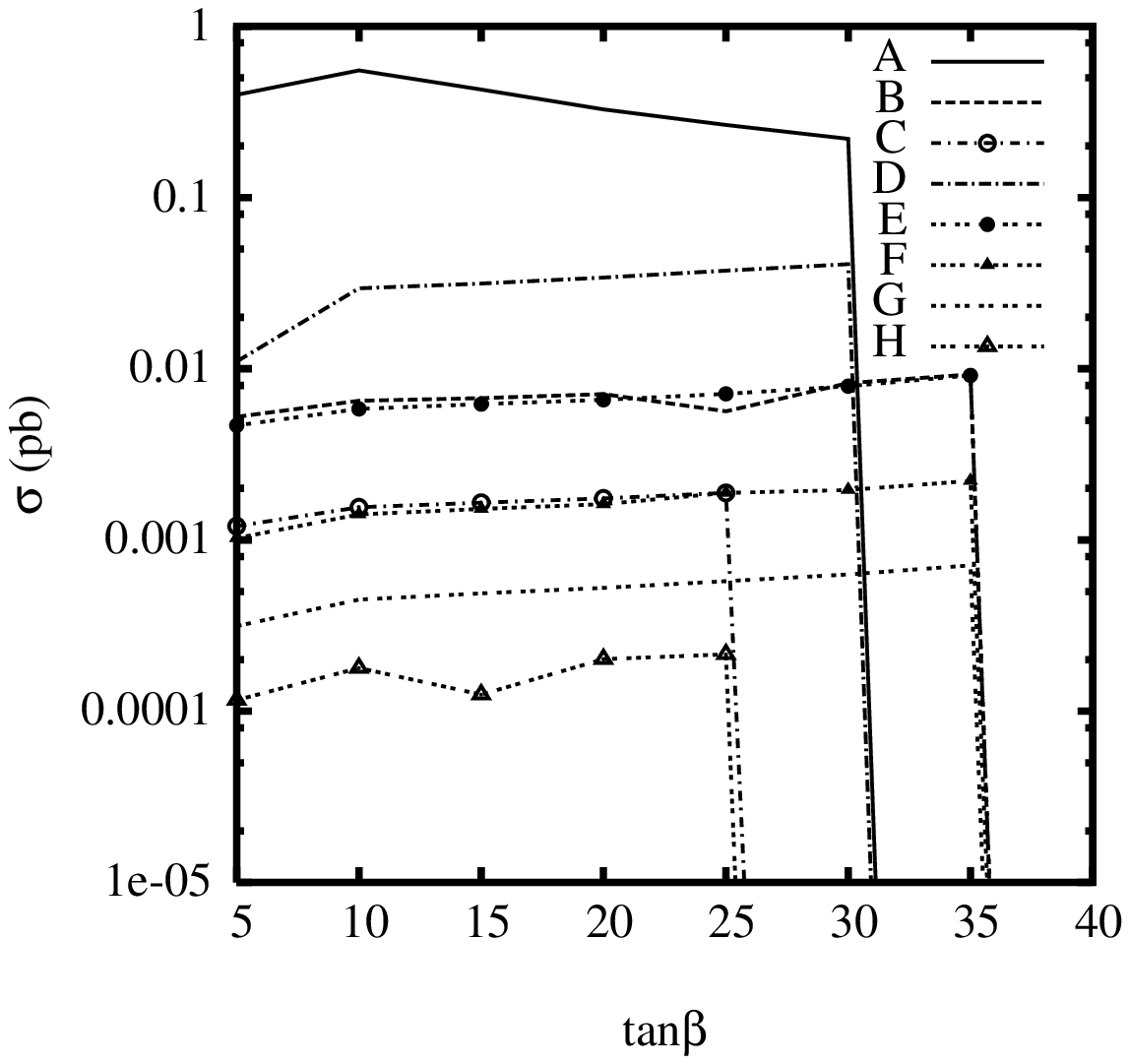}
\vspace{-7.5truecm}
\caption{Variation of the effective cross section of 
eq.~(\ref{sigma}) with $\tan\beta$ for the
 representative points in the GMSB parameter space
given in Tab. 7.}
\end{center}
\label{Fig:sigma}
\end{figure}

\clearpage\thispagestyle{empty}

\begin{figure}[t]
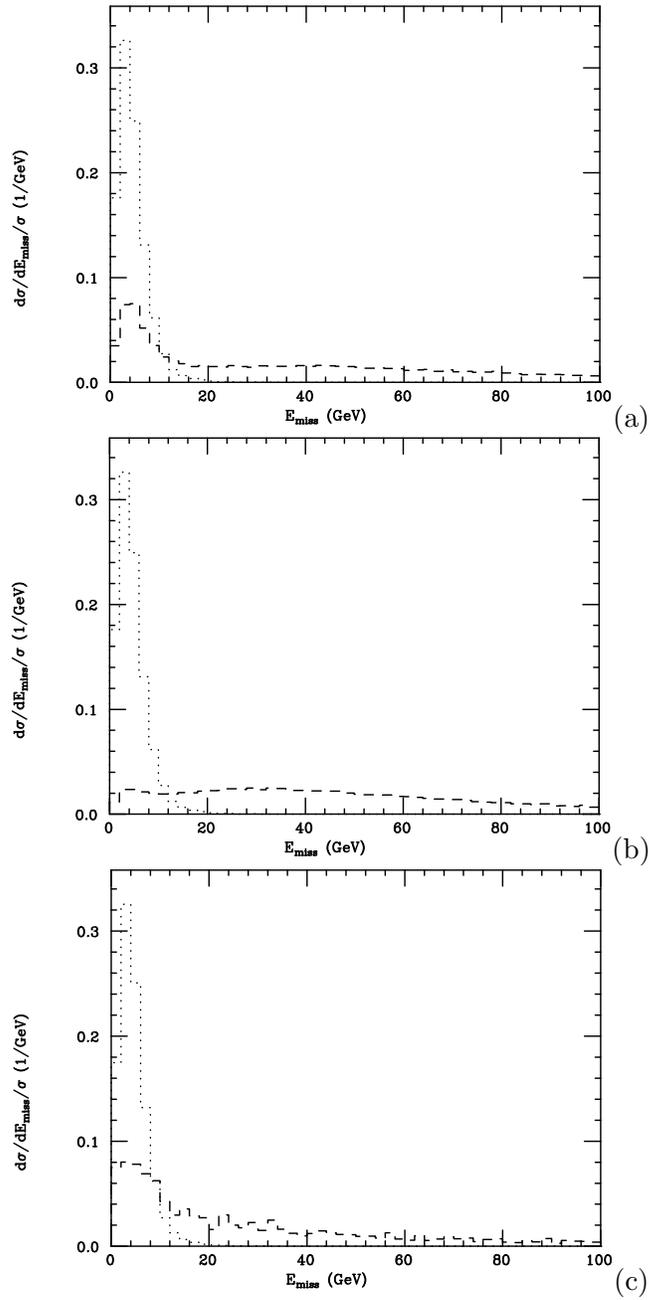

\begin{center}
\epsfig{file=newETmissA.ps,angle=90,width=8cm}(a)\\
\epsfig{file=newETmissB.ps,angle=90,width=8cm}(b)\\
\epsfig{file=newETmissC.ps,angle=90,width=8cm}(c)\\
\caption{Differential distributions in $E_{\rm{miss}}$ (as defined in
the text) normalised to one, after the cuts 1.--2. described in 
Sect.~\ref{Sect:Calculation},
as obtained from GMSB set 1 (top), 2 (middle)
and 3 (bottom) defined in Tab.~8.
The dashed(dotted) line represents the
signal(background) rates. Bins are 2 GeV wide. }
\end{center}
\label{Fig:ETmiss}
\end{figure}

\clearpage\thispagestyle{empty}

\begin{figure}[t]
\begin{center}
\epsfig{file=newmphphA.ps,angle=90,width=8cm}(a)\\
\epsfig{file=newmphphB.ps,angle=90,width=8cm}(b)\\
\epsfig{file=newmphphC.ps,angle=90,width=8cm}(c)\\
\caption{Differential distributions in $m_{\gamma\gamma}$ (as defined in
the text) normalised to pb, after the cuts 1.--2. described in 
Sect.~\ref{Sect:Calculation} and $E_{\rm miss}>20$ GeV,
as obtained from GMSB set 1 (top), 2 (middle)
and 3 (bottom) defined in Tab.~8.
The dotted(solid) line represents the
background(signal+background) rates. Bins are 3 GeV wide.}
\end{center}
\label{Fig:mphph}
\end{figure}

\clearpage\thispagestyle{empty}

\begin{figure}[t]
\begin{center}
\epsfig{file=newMphphA.ps,angle=90,width=8cm}(a)\\
\epsfig{file=newMphphB.ps,angle=90,width=8cm}(b)\\
\epsfig{file=newMphphC.ps,angle=90,width=8cm}(c)\\
\caption{Differential distributions in $M_{\gamma\gamma}$ (as defined in
the text) normalised to pb, after the cuts 1.--2. described in 
Sect.~\ref{Sect:Calculation} and $E_{\rm miss}>20$ GeV,
as obtained from GMSB set 1 (top), 2 (middle)
and 3 (bottom) defined in Tab.~8.
The dotted(solid) line represents the
background(signal+background) rates. Bins are 3 GeV wide.}
\end{center}
\label{Fig:Mphph}
\end{figure}

\end{document}